\documentstyle[11pt,newpasp,twoside]{article}
\def\edcomment#1{\iffalse\marginpar{\raggedright\sl#1\/}\else\relax\fi}
\reversemarginpar

\begin{document}
\title{The $\eta$ Chamaeleontis cluster: recent results}
\author{Warrick A. Lawson}
\affil{School of Physics, University College UNSW, Australian Defence
Force Academy, Canberra ACT 2600, Australia}

\begin{abstract}
The $\eta$ Chamaeleontis star cluster is one of the most important
nearby groupings of pre-main sequence (PMS) stars.  With accurately 
known distance, compact structure and highly-coeval intermediate-aged 
population, the $\eta$ Cha cluster is an important laboratory for 
enhancing our understanding of all PMS star evolutionary issues.
In this paper, recent results concerning the stellar population are
discussed including the discovery of a rare example of a $\sim 10$
Myr-old Classical T Tauri (CTT) star.  This and other objects point
to an X-ray-faint late-type population residing amongst the 
{\it ROSAT}-detected membership.
\end{abstract}

\section{Characterization of a new star cluster}

The $\eta$ Chamaeleontis star cluster is a recently-discovered
group of $\sim$ 10 Myr-old PMS stars at a distance of only 97 pc
(Mamajek, Lawson \& Feigelson 1999a).  The cluster was discovered
following a {\it ROSAT\,} HRI pointing at a compact group of 4
{\it ROSAT\,} All-Sky Survey (RASS) sources.  The 4 RASS sources
were recovered along with 8 additional sources.  Spectroscopy of
the optical counterparts of these sources (denoted as RECX stars 
by Mamajek et al. 1999a) showed 10 of these stars had the optical 
characteristics of late-type Weak-lined T Tauri (WTT) stars, as 
evidenced by optical activity (H$\alpha$ emission with equivalent 
widths {\it EW\,} = $1-20$ \AA) and enhanced $\lambda$6708 Li I 
absorption ({\it EW\,} = $0.4-0.6$ \AA).  The optical counterparts 
of the other two X-ray sources were the bright B8 star $\eta$ Cha 
and the A7 + A8 binary RS Cha.    

The PMS nature of the RECX stars was confirmed via {\it Hipparcos\,}
and Tycho astrometry.  {\it Hipparcos\,} parallaxes and proper motions
for $\eta$ Cha and RS Cha gave a weighted parallax for the two stars
of $\pi = 10.3 \pm 0.3$ mas (distance $d = 97 \pm 3$ pc) and proper 
motions ($\mu_{\alpha}$ cos $\delta$, $\mu_{\delta}$ = $-30$, +28 
mas\,yr$^{-1}$).  Half of the late-type RECX stars (RECX 1, 9, 10, 11 
and 12) and the X-ray-quiet A5 star HD 75505 can be related to $\eta$ 
Cha and RS Cha via common parallax and/or proper motions (Mamajek 
et al. 1999a, Mamajek, Lawson \& Feigelson 2000, Lawson et al. 2001a).  
Preliminary placement of the RECX stars on the HR diagram suggested 
a common age of $\sim$ 10 Myr (Mamajek et al.  1999a).  

Published work to date on the $\eta$ Cha cluster includes announcement 
of its discovery and analysis of spectroscopy of the RECX stars  
(Mamajek et al. 1999a), a search for microwave radio emission from 
the RECX stars (Mamajek, Lawson \& Feigelson 1999b), consideration
of the kinematic origin of the cluster and detailed analysis of the
{\it ROSAT\,} X-ray light curves (Mamajek et al. 2000; also see 
Mamajek \& Feigelson, these proceedings), and long-term photometric 
study of the late-type stars for rotation period analysis and accurate 
HR diagram placement (Lawson et al. 2001a).  In addition, Lawson \& 
Feigelson (2001) compared the cluster HR diagram to several contemporary 
sets of PMS evolutionary tracks, concluding that most models failed to 
achieve coevality across the observed range of stellar spectral types.

In this paper, some of the more recent work and forthcoming detailed 
studies of the cluster's stellar population are highlighted.

\section{Recent work and new results}

\subsection{Photometric study of the RECX stars}

Multi-epoch $V$-band photometric monitoring during 1999 and 2000 
of the 10 late-type  RECX stars showed that all were variable in 
one or both years, with the light variations ascribed to rotational 
modulation by cool starspots (Lawson et al. 2001a).  Rotation 
periods ranging from $1.3-20$ d were observed in the RECX stars, 
with the observed periods being generally consistent from 1999 to 
2000.   Some stars showed significant changes in the light curve 
amplitude between the two observing seasons, as would be expected 
from the evolution of the obscuring spots over the course of a 
year.

Calibrated {\it VRI\,} photometry allowed us to better place the 
RECX stars on the HR diagram and to infer intrinsic properties such 
as radius and temperature.  Surprisingly, comparison between the 
rotational and X-ray properties of these objects indicated the 
`saturation' level observed across many samples of PMS stars, 
log ($\it L_{\rm X}$/$L_{\rm bol}$) $\approx -3$, persists across 
all rotation periods in the RECX stars.  This unexpected finding 
strongly hints that the dynamo mechanism may be inadequate in
explaining X-ray emission in young stars!

\subsection{$\eta$ Cha -- coeval cluster}

HR diagram placement of the RECX stars (Lawson at al. 2001a)
showed a several Myr scatter in apparent age that was ascribed
significantly to unresolved binaries.  In particular, two stars
(RECX 9 and 12) appeared elevated by a factor of 2 in luminosity
above other RECX stars of similar spectral type, and there was
nearly a factor of 2 scatter in the luminosities of the mid-K
stars (RECX 1, 7 and 11).  The location of RECX 9 and 12 was
interpreted as representing the locus of near-equal-mass
binaries in the sample.

Recently, Koehler (these proceedings) observed the RECX stars 
using the ADONIS adaptive optics system on the ESO 3.6-m 
telescope and found RECX 1 and 9 to be binaries with $\approx 0.2$ 
arcsec separation (20 AU projected distance).  The $K$-band
brightness ratios were $\approx$ 1:1 (RECX 1) and 2:1 (RECX 9).
Lawson et al.  (2001a) found RECX 12 to be multi-periodic with 
1.3 and 8.6 d periods present in the light curve during both 
1999 and 2000.  Since a several Myr-old close binary can be 
tidally-locked, then one of these periods might also be the 
orbital period.  If this interpretation is correct, then the 
inferred separation of the RECX 12 binary is $< 0.25$ AU.  When
the known and probable binaries are accounted-for, cluster members 
appear highly coeval to $\pm 1$ Myr depending on the adopted 
PMS evolutionary tracks.  

Lawson \& Feigelson (2001) compared the cluster HR diagram to 
several contemporary grids of PMS evolutionary tracks, and found 
factors of $2-4$ disagreement in the inferred age and mass of 
individual cluster members.  An important test for the PMS models 
is the RS Cha binary.  Current models gave correct masses for the
binary components (compared to dynamical masses), but the inferred 
age of the binary system differed by a factor of 2.  The models 
of Seiss, Dufour \& Forestini (2000) give more self-consistent 
ages across the cluster HR diagram and color-mag diagram than 
the other models examined.

With known members ranging from spectral type B8 to M4, and
straddling a factor $> 4$ in temperature, $> 10$ in mass and
$> 1500$ in luminosity, the $\eta$ Cha cluster is thus a critical 
and difficult test for any PMS evolutionary model.

\subsection{New members - a CTT star in the $\eta$ Cha cluster}

Compared to most field stars of similar color (or temperature),
the $\eta$ Cha stars are elevated in magnitude (or luminosity) 
in the color-mag (or HR) diagram due to a combination of youth,
proximity, and low interstellar reddening towards the cluster
[Westin (1985) found $E$($b-y$) = --0.004 for $\eta$ Cha].
Also, the late-type RECX stars form a predicatable sequence in
the colour-mag (or HR) diagram due to their coevality (Lawson
et al. 2001a) and the cluster is compact ($\sim 1$ pc extent)
indicating there will be a minimal ($\sim 1$\%) geometric 
contribution to the photometry.

These favourable properties can be used to photometrically-select 
new candidate cluster members.  Follow-up spectroscopic study 
can then be conducted for signs of stellar youth, most importantly 
the presence of enhanced lithium.  Thus new members of the $\eta$
Cha cluster can be found irrespective of their X-ray properties 
and without prior knowledge of their distances or proper motions.  
(X-ray and parallax/proper motions have been used to find members
of other nearby associations such as the TW Hya association; see
e.g. Webb et al. 1999).

A limited-area search near the central regions of the cluster 
has been made by Lawson et al. (2001b), who found several 
candidate members with a similar locus in the color-mag diagram 
as the known members, of which two were subsequently identified 
as optically-active, lithium-rich low-mass objects.  The most 
significant of these is ECHA J0843.3--7905 (= {\it IRAS\,} 
F08450--7854; Moshir et al.  1989), a $V = 14.0$ CTT star with 
an optical spectrum dominated by strong Balmer and Ca II emission.
At a distance of 97 pc, the star is probably the second-closest 
known CTT star to Earth after TW Hya (if the CTT population of 
the MBM12 cloud is $> 100$ pc distant as suggested by Luhman, 
these proceedings) and another rare example of an `old' CTT 
star.  Compared to TW Hya, ECHA J0843.3--7905 is of slightly 
later spectral type (M2, compared to K7) and may have slightly 
lower H$\alpha$ activity ($EW$ = --110 \AA, compared to --140 \AA) 
although it has currently unknown H$\alpha$ variability.  Like 
other CTT stars ECHA J0843.3--7905 has a strong infrared excess, 
with a $V-$[12] excess of $\sim 5$ mag, and shows high optical 
($\Delta V = 0.6$ mag) photometric variability.

The other new cluster member is the M4 WTT star ECHA J0841.5--7853,
which is currently the lowest mass ($\approx$ 0.2 $M_{\odot}$) and 
optically faintest primary ($V = 17.1$) known in the cluster.  
Neither of these stars were detected by {\it ROSAT\,} HRI (Mamajek 
et al. 1999a, 2000), indicating we are beginning to detect a 
population of X-ray-faint low-mass cluster members with 
log $L_{\rm X} < 28.5$ erg\,s$^{-1}$; the flux of 
the weakest {\it ROSAT\,} HRI source, RECX 9.  We are also 
studying a number of other low-mass cluster candidates.

\subsection{The disk fraction for $\eta$ Cha stars}

Lyo et al. (2001) will present $L$-band photometry of the 
$\eta$ Cha stars obtained during the 1999 southern winter with 
the SPIREX telescope sited at the South Pole.  Two of the 
$\eta$ Cha stars show strong $L$-band excesses; the CTT star
ECHA J0843.3--7905 discussed above, and RECX 11.  The latter 
star was identified by Mamajek et al. (2000) as a transition
class II/III young stellar object between the CTT and WTT
phases.  RECX 11 has only moderate optical activity 
(H$\alpha\,EW = -7$ \AA) but is listed in the {\it IRAS\,} 
Faint Source Catalogue (= {\it IRAS\,} F0847--7848).  Other
RECX stars show weaker $L$-band excesses.

\section{The significance of the $\eta$ Cha cluster}

The $\eta$ Cha cluster is a nearby ($d = 97$ pc), compact 
(extent $\sim 1$ pc), unreddened, coeval 
intermediate-aged ($\sim 10$ Myr-old) grouping of PMS stars.  
This unusual combination of characteristics makes it an ideal 
laboratory for the study of all PMS evolutionary issues, e.g.
study of rotational evolution, disk dissipation and planet 
formation, magnetic activity evolution, the nature of PMS 
brown dwarfs expected to accompany the stellar population, 
the dynamical evolution of young clusters, etc.  The cluster 
has already proven to be a demanding calibrator for PMS 
evolutionary models.

\acknowledgments

WAL thanks the conference organisers for their invitation to 
contribute to this meeting.  He thanks Eric Feigelson and Eric 
Mamajek for their valuable collaboration, and Lisa Crause and 
A-Ran Lyo for their on-going significant contributions to the
study of the $\eta$ Cha cluster.  WAL's research is supported 
in-part by the Australian Research Council and University 
College Special Research Grants.

\end{document}